\documentclass[12pt]{article}
\newcommand\be{\begin{equation}}
\newcommand\ee{\end{equation}}
\newcommand\ba{\begin{eqnarray}}
\newcommand\ea{\end{eqnarray}}\newcommand\eq{\begin{equation}}           
\newcommand\en{\end{equation}}

\input epsf

\usepackage{ulem}
\usepackage{amssymb}
\usepackage{amsmath}
\usepackage{bm}
\usepackage{graphicx}
\usepackage{color}
\usepackage{subfigure}

\def\gsim{\;\rlap{\lower 2.5pt
 \hbox{$\sim$}}\raise 1.5pt\hbox{$>$}\;}
\def\lsim{\;\rlap{\lower 2.5pt
 \hbox{$\sim$}}\raise 1.5pt\hbox{$<$}\;}

\usepackage{graphicx}
\usepackage{url}
\usepackage{amssymb,amsmath}
\usepackage{amsbsy}

\begin{document}
\title{
%{\hfill  \small\\ ~\\~\\}
Cosmologically probing ultra-light particle dark matter using 21 cm signals}
 \author{Kenji Kadota$^1$, Yi Mao$^{2,3}$, Kiyomoto Ichiki$^4$ and Joseph Silk$^{2,5,6}$\\
{ \small $^1$ \it   Department of Physics, Nagoya University, Nagoya 464-8602, Japan} \\
{ \small $^2$ \it Institut d'Astrophysique de Paris, CNRS, UPMC Univ Paris 06, } \\ 
{ \small \it UMR7095, 98 bis, boulevard Arago, F-75014, Paris, France} \\
 { \small $^3$ \it Institut Lagrange de Paris, Sorbonne Universit\'es, 98 bis, boulevard Arago, F-75014 Paris, France} \\
{ \small $^4$ \it Kobayashi-Maskawa Institute for the Origin of Particles and the Universe,}\\
{ \small  \it Nagoya University,  Nagoya 464-8602, Japan}\\
 { \small $^5$  \it The Johns Hopkins University, Department of Physics and Astronomy, Baltimore, Maryland 21218, USA}\\
 { \small $^6$  \it Beecroft Institute of Particle Astrophysics and Cosmology, University of Oxford, Oxford OX1 3RH, UK}
}
\date{\vspace{-5ex}}
%\date{}  % Toggle commenting to test
\maketitle   

\begin{abstract}
%% Ubiquitous ultra-light scalar fields arise naturally in the early Universe, such as axions and more generally axion-like particles (ULPs). We develop a model-independent analysis to forecast the constraints on their mass and abundance using futuristic but realistic 21 cm observables as well as CMB fluctuations, including CMB lensing measurements. 
There can arise ubiquitous ultra-light scalar fields in the Universe, such as the pseudo-Goldstone bosons from the spontaneous breaking of an approximate symmetry, which can make a partial contribution to the dark matter and affect the large scale structure of the Universe. While the properties of those ultra-light dark matter are heavily model dependent and can vary in a wide range, we develop a model-independent analysis
to forecast the constraints on their mass and abundance using futuristic but realistic 21 cm
observables as well as CMB ﬂuctuations, including CMB lensing measurements. Avoiding the highly nonlinear regime, the 21 cm emission line spectra are most sensitive to the ultra-light dark matter with mass $m\sim 10^{-26}$ eV for which the precision attainable on  mass and abundance bounds can be of order of a few percent. 
 
\end{abstract}

\setcounter{footnote}{0} 
\setcounter{page}{1}\setcounter{section}{0} \setcounter{subsection}{0}
\setcounter{subsubsection}{0}

\section{Introduction}
The existence of light scalar fields has been explored from both particle phenomenology and cosmological aspects. A common example is the proposal of the QCD axion to solve the strong CP problem and there also have been growing interests in string axions in the so-called string axiverse scenarios \cite{pec,wein,wil,john1,ed,sv,acha}. An astrophysics example includes dark matter with $m \sim 10^{-22}$ eV, dubbed `fuzzy dark matter', which can suppress kpc scale substructure in dark matter halos because the matter cannot cluster within the Jeans scale \cite{fuz,sik,wayne4,kol,joe3} \footnote{Note that such ULPs have a Compton wavelength of order ${\cal O}(1)$ pc while the inter-particle distance has to be of order ${\cal O}(10^{-10})$ cm to contribute to the local dark matter abundance. We hence need consider their wave-like nature rather than a classical particle picture, and the Jeans scale  can here be interpreted as the de Broglie scale where the uncertainty principle prevents the localization of the ULPs.}. Another interesting parameter range for those ultra-light particles (ULPs) lies in when their mass is of order of the current Hubble scale $H_0 \sim 10^{-33}$ eV  and they play a role similar to inhomogeneous dark energy \cite{hall,amen}.
In view of the large  range of possible parameters for these light scalars, such as their mass and abundance, it would be of great interest to narrow down the allowed model parameter space for the cosmological observables in a model-independent manner.

Those ultra-light scalar fields can imprint the characteristic features on the matter power spectrum due to  `free-streaming' similar to that due to massive neutrinos \cite{kev4,les,wan,sai,bir,bat,takeu}. A wide range of the possible masses and hence the possibility for a wide range of the suppression scale in the matter power spectrum can open up a promising cosmological window to signal the existence of ULPs \cite{david3,arv,rin,pedro,davi,gra}. We characterize the ULPs by two free parameters, their mass and abundance, and we forecast future cosmological constraints on ULPs using the CMB, including  CMB lensing, and 21 cm observables \cite{lewrep,furrep} 
\footnote{Even though there are a wide range of possibilities for the interactions of the ULPs, here we just need to consider the gravitational interactions in forming the large scale structure of the Universe. The ULPs in this paper simply refer to a dark matter component with ultra-light mass $\ll 1$eV which can be as light as the current Hubble scale $H_0 \sim 10^{-33}$eV.}.
%% \footnote{Even though there are a wide range of possibilities for the interactions of the ULPs (for instance, in contrast to the conventional axions, the coupling to the gluon field strength can be absent for the ULPs in the string setup \cite{john1,ed,acha}), here we just need to consider the gravitational interactions in forming the large scale structure of the Universe. The ULPs in this paper simply refer to a dark matter component with ultra-light mass $\ll 1$eV which can be as light as the current Hubble scale $H_0 \sim 10^{-33}$eV.}.
Because of a large number of modes available for observing the high redshift matter distribution along with redshift information, the 21 cm line of neutral hydrogen possesses promising power for probing the matter power spectrum with unprecedented precision \cite{furrep}. We aim to clarify the range of the mass and abundance of ULPs making up part of the total matter of the Universe (in addition to the dominant ordinary cold dark matter (CDM)) which  future 21 cm observables can probe.  

The paper is organized as follows. \S \ref{sup} outlines the effects of the ULPs on the matter power spectrum that will be probed by 21 cm signals. \S \ref{fishres} gives a brief review of the Fisher analysis formalism using 21 cm observables followed by the main results of our paper which present forecasts of  errors of the mass and abundance of the ULPs.

\section{Suppression in the matter power spectrum}
\label{sup}

%%%%%%%%%%%%%%%%%%%%%%%%%%%%%%%%%
\begin{figure}%[htb!]
\begin{center}    
\epsfxsize = 0.48\textwidth
%\epsffile{./mathematica/newnb/avesigmav1.eps}
%\epsffile{./mathematica/newnb/avesigmav2.eps}
%\epsffile{./nov19axpertb.eps} <- fa=0.01, inverted
\epsffile{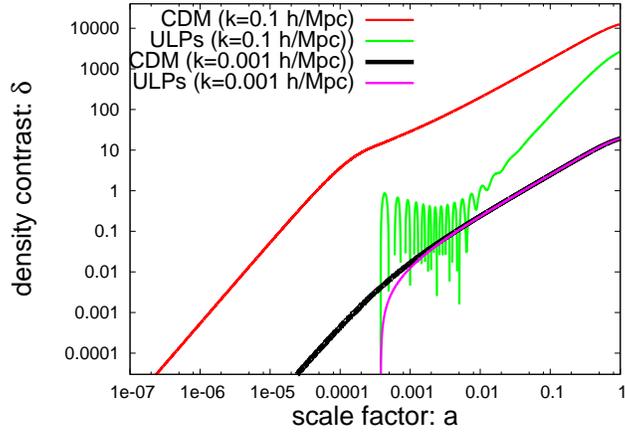}
\end{center}        
\caption{The perturbation evolutions for ULPs ($m_u=10^5H_0,f_u=0.05$) and CDM.}
\label{nov19axpertb}
\end{figure}
%%%%%%%%%%%%%%%%%%%%%%%%%%%%%%%%%

%%%%%%%%%%%%%%%%%%%%%%%%%%%%%%%%%%%%%%%%%%%%%%%
\begin{figure}[htbp]
 \begin{minipage}{0.5\hsize}
  \begin{center}
   \includegraphics[width=\linewidth]{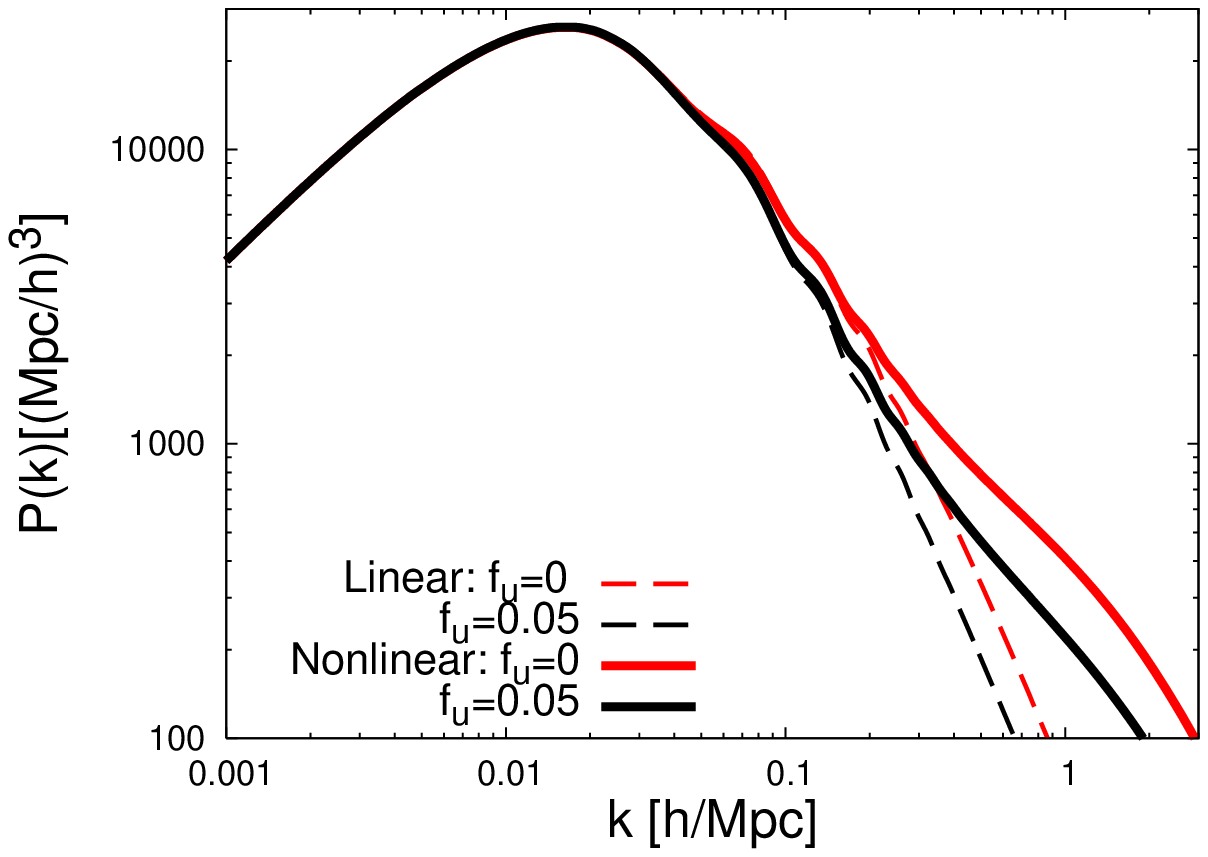}
  \end{center}
%  \caption{The perturbation evolutions for the ULP and CDM.}
%  \label{nov15axpert}
 \end{minipage}
 \begin{minipage}{0.5\hsize}
  \begin{center}
   \includegraphics[width=\linewidth]{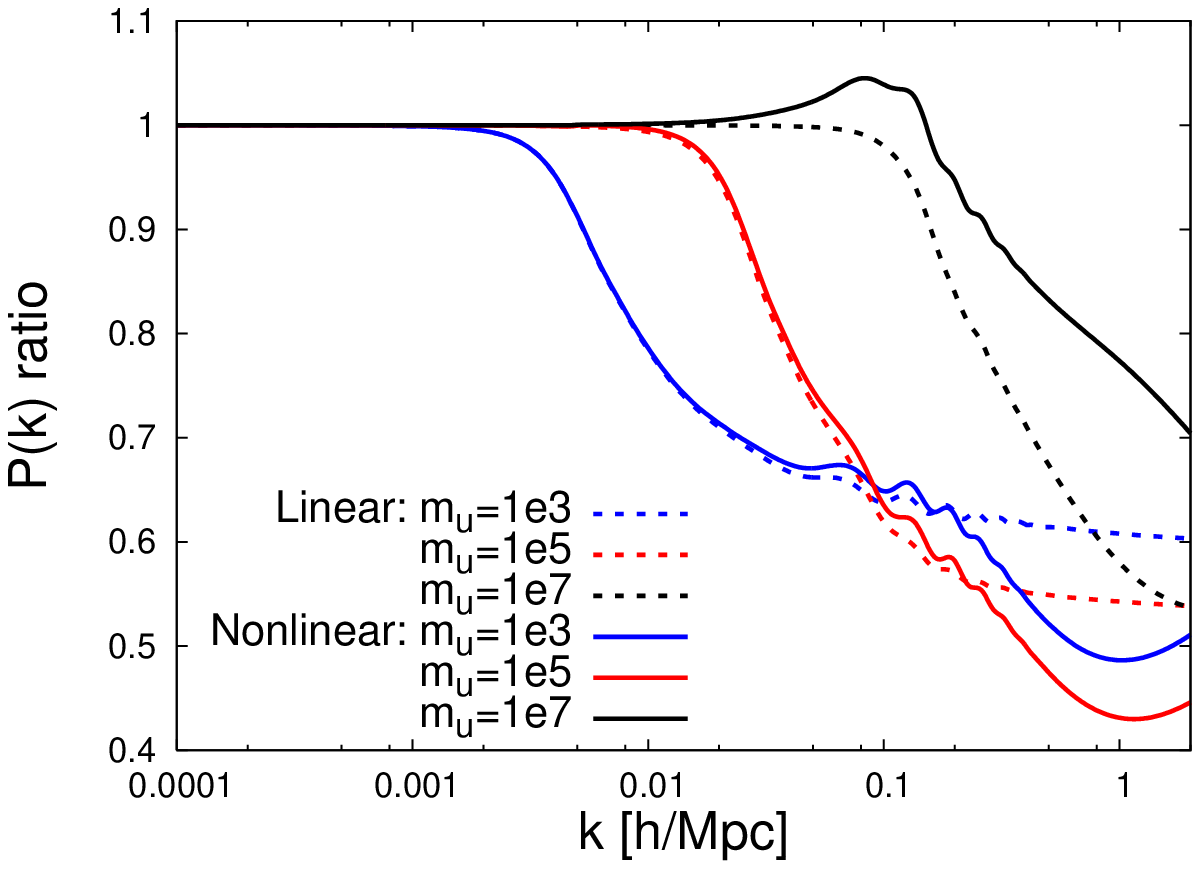}
  \end{center}
%  \caption{The transfer function $T^2(k)$}
%  \label{nov20pratiob}
 \end{minipage}
\caption{Left: The (linear and nonlinear) power spectrum $P(k)$ with and without the ULPs for $m_u=10^5 H_0, f_u=0.05$. Right: The transfer function $T^2(k)=P(k)_{\rm{ULPs}}/P(k)_{\rm{no~ULPs}}$ representing the ratio of the power spectrum including the ULPs ($f_u=0.05$) to that without the ULPs (the values of $m_u$ in the figure are in terms of $H_0\approx 2\times 10^{-33}eV$).}
\label{pertfig}
\end{figure}
%%%%%%%%%%%%%%%%%%%%%%%%%%%%%%%%%%%%%%%%
Let us briefly review the conventional Jeans analysis in order to clarify the characteristic features for the relativistic species imprinted in the matter power suppression. The conventional Jean analysis tells us that the leading order perturbation equation possesses the gravitationally stable solution for a short wave-length mode $k\gg k_J$ and the unstable (growing) one for $k \ll k_J $ ($k_J$ is the Jeans wave number) \footnote{The Klein-Gordon equation  that the ULPs obey can be mapped to the continuity and Euler equations for a relativistic fluid and the familiar Jeans analysis noting the pressure term in the Euler equation follows \cite{wayne4}.}. The pressure inside the Jeans scale prevents the matter from collapsing and, in contrast to the conventional matter growth proportional to the linear growth factor $\delta \propto D(z)$ after matter-radiation equality, the matter density perturbations grow more slowly as $\delta \propto (1-f) D(z)^{1-p}$ when there exists a fraction $f$ of the matter component which does not cluster due to the pressure support ($p=(5-\sqrt{25-24 f })/4$) \cite{john1,fuz,wayne4,bond}. A notable feature for the ULPs is its effective sound speed which is scale-dependent and can be less than unity for a large scale factor $a$ ($c_s \approx k/2m_ua$ for $a \gg k/2m_u$, and $c_s \approx 1$ below the Compton scale $a \ll k/2m_u$ ($m_u$ denotes the ULP mass)) \cite{fuz,wayne4}. 
%% More quantitatively, for the leading order perturbation equation 
%% \ba
%% \ddot{\delta_k}+2 H \dot{\delta_k}+\left( \frac{c_s^2 k^2}{a^2}-4\pi G \rho_m  \right) \delta_k=0
%% \ea 
%% there exists a gravitationally stable solution for short wave-length mode $k\gg k_J$ and an unstable (growing) one for $k \ll k_J $ with the Jeans wave number which can be obtained by equating the pressure term $(k^2/a^2)c_s^2 \delta_k$ and the source term $4\pi G \rho_m \delta_k$ ($\rho_m$ is the matter density).
The Jeans wave number for the ULPs with the effective sound speed then becomes $k_J(a)=2a(\pi G \rho_m(a))^{1/4} m_u^{1/2}$  for $a \gg k/2m_u$. We consider  scenarios in this paper where the ULP behaves like dark energy due to the large Hubble friction  for $H>m_u$ and starts oscillations, behaving like dark matter, once $H \leq m_u$. We implemented ULPs into CAMB \cite{camb} such that the ULPs follow the cosmological constant-like equation of state $w\equiv P/\rho=-1$ for $H(t)>m_u$ and the matter-like equation of state $w=0$ for $H(t) \leq m_u$ \footnote{We assume the sudden transition in the ULP equation of state at $m=H(t)$ which suffices for our purpose of illustrating the cosmological power of constraining the ULP properties. The detailed treatment of this transition keeping track of the slowly rolling regime and rapidly oscillating equation of state before the system settles down is beyond the scope of this paper because of too wide a time scale between the cosmic expansion and scalar field dynamics. We refer the reader to, for instance, Ref. \cite{david3} for studies of this transition period.}. The evolution of the ULP fluctuations $\delta_u=\delta \rho_u/\rho_u$ is shown in Fig. \ref{nov19axpertb} for the ULP mass and fraction $m_u=10^5 H_0 (H_0\approx 2 \times 10^{-33}$eV$), f_u=0.05$ ($f_i=\frac{\Omega_i}{\Omega_m}$ represents the fraction of the matter species $i$ with respect to the total matter $\Omega_m=\Omega_{b}+\Omega_d=\Omega_{b}+\Omega_{cdm}+ \Omega_u+\Omega_{\nu}$ (representing, respectively, the baryon, cold dark matter, ULPs and neutrinos. $\Omega$ represents the fraction with respect to the critical density)). The fluctuations represented in Fourier space $\delta(k)$ cannot grow when they behave like a cosmological constant and can start growing once the ULPs start to oscillate. The perturbation growth however is suppressed inside the Jeans scale and the perturbation growth has to wait till it goes outside the Jeans scale for a large enough  value of $a$. We also plotted the CDM perturbation evolution which illustrates  that the ULP perturbations can catch up with the CDM perturbations for small $k$ but not for large $k$, analogously to the familiar behavior of the baryon perturbation evolution. The nonlinearity becomes important when $k^3 P(k) /(2\pi^2)$ becomes of order unity. We calculate the power spectrum from CAMB, as shown in the left-hand side of Fig. \ref{pertfig}, and use  Halofit \cite{smi,ryo} to map the linear power spectrum including the ULPs and neutrinos to the nonlinear one \footnote{This mapping by Halofit from the linear to the nonlinear matter power spectrum can be affected by ULPs and neutrinos. The modification of the Halofit fitting formula by taking account of these light species is however beyond the scope of our paper and we refer the reader to, for instance, Ref. \cite{sai,bir} for more details on the impact of those light species on the nonlinear matter power spectrum.}. The nonlinearity becomes important for $k \gtrsim 0.1 [h/$Mpc], where the deviation between the linear and nonlinear treatments becomes large. In our parameterization, increasing $f_u$, while keeping $\Omega_{m}h^2$ and $\Omega_b h^2$, enhances the baryon to cold dark matter ratio and the nonlinear power spectrum captures these enhanced baryon oscillation effects.

We can analytically estimate that, if the oscillations start during the matter domination epoch, the moment of oscillation is around
\ba
\label{matzoci}
z_{osc} \sim \left( \frac{m_u^2}{H_0^2 \Omega_m} \right)^{1/3} \sim 1.5 \left( \frac{m_u}{H_0}\right)^{2/3} 
\left(\frac {0.14}{\Omega_m h^2}\right)^{1/3}
\ea
%before which the ULP behaves as the dark energy with a constant $\rho_{a}$ and after which as the matter $\rho_{a} \propto 1/a^3$.
If the oscillations start during the radiation-dominated epoch, 
\ba
\label{radzosci}
z_{osc} \sim \frac{m_u^{1/2}z_{eq}^{1/4}}{H_0^{1/2} \Omega_m ^{1/4}} \sim 10  \left(\frac{m_u}{H_0}\right)^{1/2}
\left(\frac{z_{eq}}{3200}\right)^{1/4} \left(\frac{0.14}{\Omega_m h^2}\right)^{1/4}
\ea
We can hence estimate that $m_u\sim 10^5  H_0$ leads to the oscillation starting around the matter-radiation equality epoch $z_{osc}\sim 3200 (\sim z_{eq})$ \footnote{This turns out to give the right order of magnitude for our numerical evaluation of $m_u\sim 1.4\times 10^5 H_0$ for $z_{osc}\sim z_{eq}$ when $f_u=0.05$. Another characteristic scale $z_{osc}\sim 1100 (\sim z_{CMB})$ corresponds to $m_u \sim 2.3\times 10^4H_0$.}.  
For the modes which enter the horizon during matter domination, we can analytically estimate that the suppression in the matter power spectrum starts around the scale corresponding to the Jeans scale when the ULP starts oscillating $k \sim \left(H_0^2 \Omega_m  \right)^{1/3}m_u^{1/3}$. % \sim 0.008 Mpc^{-1} \left( \left(\frac{m_u}{ 10^{-30}eV}\right) \left( \frac{\Omega_m h^2 }{0.12}\right) \right)^{1/3}
%\ea 
Similarly, when the oscillations start during radiation domination, the suppression is expected to occur for scales smaller than the Jeans scale at matter-radiation equality $k \sim (m_u^2 H_0^2 \Omega_m a_{eq})^{1/4}$. The suppression scales for different masses are illustrated in the right-hand side of Fig. \ref{pertfig} which shows the transfer function $T^2(k)=P(k)_{\rm{ULPs}}/P(k)_{\rm{no~ULPs}}$ representing the ratio of the power spectrum including the ULPs to that without ULPs. We are particularly interested in the ULP masses which affect the matter power at the 21 cm-observable scales of $0.055 \lesssim k\lesssim  0.15 [$Mpc$^{-1}]$. We can see that the baryon acoustic oscillation effects are more prominent in the nonlinear matter power spectrum than in  the linear one \cite{mei,seo,cro,taruya4,jeon} and $m_u \sim 10^7H_0$ lets the suppression start right in the 21 cm observable range \footnote{We also note that the asymptotic value of transfer function differs for a different mass even with the same $f_U$, because the period of the perturbation growth during which the free streaming suppression is relevant is different for a different mass \cite{amen,pedro,eh97}}. 
\section{Forecasts}
\label{fishres}
\subsection{Formalism}
To forecast the constraints on the cosmological parameters including those relevant to the ULPs, we perform the Fisher likelihood analysis for future 21 cm experiments. We also use the CMB observables including CMB lensing which help remove the parameter degeneracies  that the $21$ cm signals would otherwise suffer from. We briefly outline the formalism of the likelihood analysis here, and present the results in the next subsection. 

We first review the 21 cm Fisher analysis\footnote{We refer the readers to Ref. \cite{furrep} and references therein for more details on the 21 cm physics}. The 21 cm radiation comes from the atomic transition between the two hyperfine levels of the hydrogen $1s$ ground state. In the linear regime, the power spectrum of 21~cm brightness temperature fluctuations can be written as  
%($\mu_{{\bf k}} \equiv {\bf k} \cdot {\bf n}/|{\bf k}|$ is the cosine fo the angle between the line of sight and the Fourier mode wave vector ${\bf k}$ )
%($\mu_{{\bf k}}$ is the cosine of the angle between the line of sight and the comoving wave vector ${\bf k}$ )
\ba
 P_{\Delta T}({\bf k},z)=\widetilde{\delta T_b}^2 \overline{x}_{H_{I}}^2 [b_{ H_{I}}(z)+\mu_{{\bf k}}^2]^2~ P_{\delta \delta}(k,z)
%P_{\Delta T}=\overline{T}_b^2 \overline{x}_{H_{I}}^2 [b_{ H_{I}}(k,z)+\mu_{{\bf k}}^2]^2~ P_{\delta \delta} 
\label{21peq}
\ea
 where $\mu_{{\bf k}}$ is the cosine of the angle between the line of sight ${\bf n}$ and the comoving wave vector ${\bf k}$. $P_{\delta \delta}$ is the total matter fluctuation and we assume the baryon density distribution follows that of the total matter $\delta_{\rho}=\delta_{\rho_H}$. $\widetilde{\delta T_b}(z)=(23.88 {\rm mK}) \left( \frac{\Omega_b h^2}{0.02}\right)  \sqrt{ \frac{0.15}{\Omega_m h^2} \frac{1+z}{10}}$. Here we consider $z \lesssim 10$ when the spin temperature $T_S \gg T_{CMB}$, so that the dependence of 21 cm brightness temperature on $T_S$ drops out.\footnote{
$T_s\gg T_{CMB}$ can be justified soon after the reionization begins because the gas temperature can be much higher than the CMB temperature due to the heating of  the IGM to hundreds of Kelvin by the X-ray background from the first stars, and a large number of Ly$\alpha$ photons from star formations can couple the spin temperature to the gas temperature. This helps to reduce the potentially large uncertainties in the determination of the spin temperature.} We define the neutral and ionized density bias, $b_{H_{I}}(z)$ and $b_{H_{II}}(z)$, as the ratio of the density fluctuation in the neutral hydrogen $H_{I}$ and ionized hydrogen $H_{II}$, respectively, to that of total matter density in Fourier space, i.e.~$b_{H_{I}}\equiv \delta_{\rho_{H_{I}}}(k)/\delta_{\rho}(k)$, $b_{H_{II}}\equiv \delta_{\rho_{H_{II}}}(k)/\delta_{\rho}(k)$. They are related by $b_{H_{I}}=(1-\overline{x}_{H_{II}} b_{H_{II}})/\overline{x}_{H_{I}}$, where the global neutral and ionized fractions are related as $\overline{x}_{H_{I}}+\overline{x}_{H_{II}}=1$. We use the excursion set model of reionization \cite{esmr} to obtain the fiducial values of ionized density bias $b_{H_{II}}(z)$ and the mean ionized fraction $\overline{x}_{H_{II}}(z)$. The actual radio interferometric arrays measure the 21 cm signals from coordinate ${\bf \Theta} \equiv \theta_x \hat{e}_x+ \theta_y \hat{e}_y+\Delta f  {\bf n}$, where $(\theta_x,\theta_y)$ represent the angular location on the sky plane and $\Delta f$ is the frequency difference from the central redshift $z_*$ of a redshift bin. The Fourier dual of ${\bf \Theta}$ is ${\bf u}\equiv u_x \hat{e}_x + u_y \hat{e}_y +u_{\parallel}{\bf n}$. Here ``$\perp$'' and ``$\parallel$'' represent the perpendicular and parallel projections to the line of sight, respectively, and $u_{\parallel}$ has units of time. ${\bf \Theta}$ and ${\bf u}$ are related to ${\bf r}$ and ${\bf k}$ by ${\bf \Theta}_{\perp}={\bf r}_{\perp}/d_A(z_*),\Delta \nu=r_{\parallel}/y(z_*)$, and ${\bf u}_{\perp} = d_A(z_*) {\bf k}_{\perp},u_{\parallel}=yk_{\parallel}$ ($d_A$ is the comoving angular diameter distance, $y(z)\equiv \lambda_{21}(1+z)^2 /H(z),\lambda_{21}=\lambda(z)/(1+z)= 21$cm). We use the actual 21 cm observable $P_{\Delta T}({\bf u})=P_{\Delta T}({\bf k})/d_A^2 y$, rather than $P_{\Delta T}({\bf k})$, in our Fisher matrix for 21 cm power spectrum measurements \cite{mc,mao}
\ba
%F^{21cm}_{\alpha\beta}=\sum_{{\bf u}}  \frac{1}{[\delta P_{21} ({\bf u})]^2}  \left( \frac{\partial P_{21}({\bf u})}{\partial p_{\alpha}}\right)  \left( \frac{\partial P_{21}({\bf u})}{\partial p_{\beta}}\right) 
F^{\rm 21cm}_{\alpha\beta}=\sum_{{\bf u}}  \frac{1}{[\delta P_{\Delta T} ({\bf u})]^2}  \left( \frac{\partial P_{\Delta T}({\bf u})}{\partial p_{\alpha}}\right)  \left( \frac{\partial P_{\Delta T}({\bf u})}{\partial p_{\beta}}\right) 
\ea
where $\{ p_{\alpha} \}$ represent the free parameters in our model. We assume a logarithmic pixelization $du_{\perp}/u_{\perp}=du_{\parallel}/u_{\parallel}=0.1$. The error in power spectrum measurement is $\delta P_{\Delta T} ({\bf u})=[P_{\Delta T}({\bf u})+P_N(u_{\perp})]/\sqrt{N_c}$, where $N_c=u_{\perp} du_{\perp} du_{\parallel}\Omega B/(2 \pi^2)$ is the number of independent modes in each pixel ($\Omega$ is a field of view solid angle and $B$ is the bandwidth of a redshift bin). $P_N$ is the noise power spectrum $P_N({\bf u}_{\perp},z)=(\lambda T_{sys}/A_e)^2 /(t_0 n({\bf u}_{\perp}))$, where $T_{sys}\approx (280{\rm K})[(1+z)/7.4]^{2.3} $ is the system temperature \cite{sys}, $A_e$ is the effective collecting area of each antenna tile, and $t_0$ is the total observation time. We assume the interferometric arrays have antennae concentrated inside a nucleus of radius $R_0$ with almost $100 \%$ coverage fraction, and the coverage density drops as $r^{-2}$ inside the core from $R_0$ to $R_{in}$. The number of antennae within $R_{in}$, $N_{in}$, and the fraction of the antennae within $R_0$, $\eta$, are related according to $R_0=\sqrt{\eta N_{in}/\rho_0 \pi},R_{in}=R_0 \exp[(1-\eta)/(2\eta)]$ where $\rho_0$ is the central array density \cite{mao}. For concreteness, we assume an Omniscope-like instrument \cite{teg1} consisting of a million 1m $\times$ 1m dipole antennae with a field of view of $2 \pi$ steradians whose specifications are $(N_{in},L_{min},\eta,A_e(z=6/8/12)[m^2],\Omega[sr])=(10^6, 1, 1, 1/1/1,2\pi)$ and we assume $t_0=4000$ hours for each redshift bin of bandwidth $B=6$MHz. We also assume the residual foregrounds can be neglected for $k_{\parallel} \geq k_{\parallel,{\rm min}}=2\pi/(yB)$ \cite{mc}, and the minimum baseline $L_{min}$ sets $k_{\perp,{\rm min}}=2\pi L_{min}/(\lambda d_A)$ (for example, for an Omniscope-like array, $k_{min}\approx k_{\parallel,{\rm min}}=0.055$/Mpc at $z=10.1$). We conservatively restrict our studies to large scale $k \leq 0.15 /$Mpc for the sake of the linear treatment of 21 cm observables, to avoid any scale-dependent bias at the nonlinear regime and the nonlinear effects due to reionization patchiness at the scale of the typical size of ionized regions \cite{Shapiro13}.

The CMB can also be affected by light dark matter through the change in matter-radiation equality and also via the Sachs-Wolfe effect. The CMB is also helpful in removing the degeneracies among the cosmological parameters. The CMB lensing is in particular helpful in removing the so-called geometric degeneracy which the primary CMB observables would otherwise suffer from \cite{dick3,mati,she,how}. We consider the CMB observables $T, E,d$ which represent the CMB temperature, polarization and  CMB deflection angle respectively \footnote{See, for instance, Ref. \cite{lewrep} for a review on CMB lensing.}. We assume the Planck-like specifications \cite{adepl2} including the CMB lensing measurements covering up to the multipole $l_{max}=2500$, three channels $100, 143, 217$ GHz and the sky coverage $f_{sky}=0.65$.

%The CMB can also be affected by the light dark matter through the change in the matter-radiation equality and also via the Sachs-Wolfe effect. The CMB is also helpful in removing the degeneracies among the cosmological parameters. The CMB lensing is in particular helpful in removing the so-called geometric degeneracy which the primary CMB observables would otherwise suffer from \cite{dick3,mati,she,how}. We consider the CMB observables $T, E,d$ which represent the CMB temperature, polarization and  CMB deflection angle respectively \footnote{See, for instance, Ref. \cite{lewrep}, for the review on the CMB lensing.}.
% We assume the Planck like specifications \cite{adepl2} including the CMB lensing measurements covering up to the multipole $l_{max}=2500$, three channels $100, 143, 217$ GHz and the sky coverage $f_{sky}=0.65$. %, and the CMB lensing statistical noise is estimated by the optimal quadratic estimator method of Hu \& Okamoto \cite{camb,pedro,take,oka2}. 
%For the 21 cm observables, we considered the Ominiscope as the fiducial future 21 cm experiments. 

The corresponding Fisher matrix is \cite{kar}
\ba
F_{\alpha \beta}^{CMB}
=\sum _{l=2}^{l_{max}}
\frac{f_{sky}(2l+1)}{2}
Tr [{\bf  C},_{\alpha} {\bf \tilde C}^{-1}{\bf C},_{\beta} {\bf \tilde C}^{-1} ]
%\frac{f_{sky}(2l+1)\Delta l}{2} Tr [D_{\alpha} \tilde{C}^{-1}D,_{\beta} \tilde{C}^{-1} ]
\ea
with the symmetric matrix ${\bf \tilde C}$ including both signal and noise given by
\ba
{\bf \tilde C}= {\bf  C}+{\bf  N}=
 \left( \begin{array}{ccccc}
C_l^{TT}+N_l^{TT} & C_l^{TE} & C_l^{T d}  \\
C_l^{TE}& C_l^{EE}+N_l^{EE} & 0 \\
C_l^{T d} & 0  & C_l^{d d}+ N_l^{d d} \
\end{array} \right)
\ea
${\bf C},_{\alpha}$ refers to the partial derivative with respect to a cosmological parameter $p_{\alpha}$.
Note the noise term $N_l$ contributes only to the auto-correlation spectra. For $N_l^{TT}$ and $N_l^{EE}$, we simply consider the dominant detector noise represented by the photon shot noise \cite{dick3,knox}, and the CMB lensing statistical noise is estimated using the optimal quadratic estimator method of Hu \& Okamoto \cite{take,oka2}. 
%Note in this covariance calculation, $C_l^{T,E,\gamma}$ etc except the CMBlensing poetntial $C^{\psi}$ are the unlensed Cell assuming no lensing (or the power spectrum of the estimator, ie estimate the original Cell by delensing the lensed spectrum, so these Cell are original one before being lensed). $C_l^{T \psi}$ is the cross correlation between the CMB lensing potential and the unlensed T (ie T before being lensed). The CMB lensing is odd parity (like B) so $E \psi$ term vanishes.
 The total Fisher matrix was obtained by adding the 21 cm and CMB Fisher matrix $F \approx F^{21 cm}+F^{CMB}$. \footnote{We did not take account of the potential cross correlation between $F^{21cm}$ and $F^{CMB}$ which is beyond the scope of this paper and we refer the readers to, for instance, Ref. \cite{koma} for the possible correlations between $F^{21cm}$ and $F^{CMB}$.} The modified version of the CAMB \cite{camb} was used to obtain the CMB and matter power spectra where the ultra-light fluid component was implemented in the Boltzmann equations. 

\subsection{Results}
\begin{figure}%[htb!]
\begin{center}    
\epsfxsize = 0.48\textwidth
%\epsffile{./mathematica/newnb/avesigmav1.eps}
%\epsffile{./mathematica/newnb/avesigmav2.eps}
\epsffile{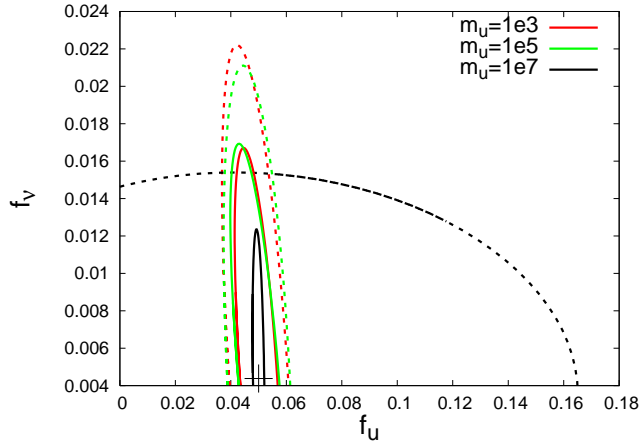}
\end{center}        
\caption{$1 \sigma$ error contour for the ULP and neutrino fractions with respect to the total matter $f_u,f_{\nu}$. The solid curves are the contours from both 21 cm and CMB observables while the dashed curves are for the CMB alone. The fiducial values $(f_u,f_{\nu})=(0.05,0.0044)$ for the normal neutrino mass hierarchy is indicated by $+$.}
\label{kenelli}
\end{figure}

Let us first clarify our conventions. We vary 12 parameters in our Fisher analysis (the numerical values in the parentheses are the fiducial values \cite{adepl})
$\Omega_{\Lambda}$ (0.69), $\Omega_m h^2$ (0.14), $\Omega_b h^2$ (0.022), $n_s$ (0.96), $A_s$ (scalar amplitude) ($2.2 \times 10^{-9}$), $\tau$ (reionization optical depth) (0.095), $N_{eff}$ (the effective number of relativistic neutrino species), $m_{a}$, $f_{a}$, $f_{\nu}$, $x_{H_{I}}(z)$, $b_{H_{II}}(z)$. % and the fiducial values, unless stated otherwise, are $\Omega_{\lambda}$, $\Omega_m h^2$, $\Omega_b h^2$, $n_s$, $A_s$, $tau$,$f_a$=$(0.6914, 0.14305,0.022161,0.9611,2.2131 \times 10^{-9},0.0952)$. 
%We take the fiducial value $f_a=0.01$
 The total matter density consists of $\Omega_m=\Omega_{b}+\Omega_d=\Omega_{b}+\Omega_{cdm}+ \Omega_u+\Omega_{\nu}$ and 
%OK. lets use the same convention.
%\ba
$\Omega_m=1-\Omega_{\Lambda}-\Omega_k$ with $f_i=\frac{\Omega_i}{\Omega_m}$. We use the reduced Hubble parameter $h=\sqrt{ \Omega_mh^2/(1-\Omega_{\Lambda})}$ to keep the flatness of the Universe $\Omega_k=0$. 
% where .  We use the reduced Hubble parameter $h=\sqrt{ \Omega_mh^2/(1-\Omega_{\Lambda})}$ to keep the flatness of the Universe $\Omega_k=0$.
% and hence the
% value of the Hubble parameter was accordingly adjusted to keep the
% flatness and ,\Omega_m=\Omega_{b}+\Omega_{CDM}+
% \Omega_{\nu}+\Omega_{a}$. 
For the fiducial models, unless stated otherwise, we use $x_{H_{I}}=0.5$ at the redshift bin of $z=10.10$ and $b_{H_{II}}=5.43$ obtained by the excursion set model of reionization \cite{esmr}, and the power spectrum up to the scale $k_{max}=0.15$/Mpc was used.
The matter power suppression features are common to the light species and the familiar example is that of the neutrino species which can worsen the ULP parameter estimations due to the parameter degeneracies. We choose the conventional normal mass hierarchy scenario for our fiducial neutrino mass pattern consisting of three neutrinos $(m_{\nu_1},m_{\nu_2},m_{\nu_3})=(0,0.009,0.05)$[eV] based on the global analysis of neutrino oscillation data giving $\Delta m_{31}^2=2.47 \times 10^{-3}$eV$^2, \Delta m_{21}^2=7.54 \times 10^{-5}$ eV$^2$ where $ \Delta m_{ij} \equiv m_i^2-m_j^2$ \cite{fogi3,pdg} (accordingly we choose $N_{eff}=1.046,f_{\nu}=0.0044$). %including the measurements of the neutrino mixing angle $\theta_{13}$ at reactor experiments
%%  \cite{fogi3} 
%% motivated by hte inverted mass hierachy sanario\ \footnote{Choosing the normal mass hierchy for the neutrino mass pattern does not affect our discussions, the difference being that the neutrino effects on the parameter estimatin becomes weaker for the normal mass hierchys cenario because the sum of the total neutrino masses becomes smaller. While the neutrino oscillation experiments  \cite{pdg,fogi3} fix the mass slittings, the lightest neutrino mass can still be free parameter. We simply took it to be massless as often conventionally done in the lietature, and refer the readers to, for instance, Ref. \cite{} for the effects of the lightest neutrino mass, in the normal and inverted mass hierrhy scanrios, on the cosmological parameter estimations.}  
Because of the similar effects to suppress the matter power, we can expect the negative correlation between $f_u$ and $f_{\nu}$. This is confirmed in Fig. \ref{kenelli} which shows the $1\sigma$ error contours with all the other parameters marginalized over, even though
there do exist the distinctive features between the ULPs and neutrinos such as the ULPs'
scale dependent effective sound speed and transition from the dark
energy to dark matter like behavior which the neutrinos do not possess.
Consequently, the precise measurements of the power spectrum around the suppression
starting scale for each species should be able to distinguish these species from one another.
%Following our observation in the last section that the sensitive mass range for the CMB observables is around $m_u \sim 10^{4\sim 5}/ H_0$ corresponding to the oscillation starting around the last scattering epoch, %while that for the 21 cm is around $m_u\sim 10^{7} H_0$ which can give the significant change in the matter power in the 21 cm observable scales, 
Fig. \ref{kenelli} indeed shows the tendency of the CMB losing the sensitivity to the ULPs for $m_u \gg 10^{5} H_0$ because the ULP oscillation starts well before the last scattering surface epoch for such a large $m_u$. The CMB observables however are still essential to improve the constraints on ULPs from the 21 cm observables because of lifting the degeneracies among the cosmological parameters. For instance, the 21 cm alone without adding the CMB observables cannot constrain the ULP parameters so well because of too strong degeneracies between $A_s$ and $x_{H_{I}}$ both of which affect the 21 cm power spectrum amplitude as given in Eq. \ref{21peq}.

%The large errors in 21cm only case is due to the large degeneracy between A_s and x_HI. Both of them are proportional to the overall amplitude of power spectrum, so effectively the 21cm power spectrum is only sensitive to the product of them. 

%%%%%%%%%%%%%%%%%%%%%%%%%%%%%%%%%
%%%%%%%%%%%%%%%%%%%%%%%%%%%%%%%%%

%%%%%%%%%%%%%%%%%%%%%%%%%%%%%%%%%%%%%%%%
\begin{figure}[htbp]
 \begin{minipage}{0.5\hsize}
  \begin{center}
   \includegraphics[width=\linewidth]{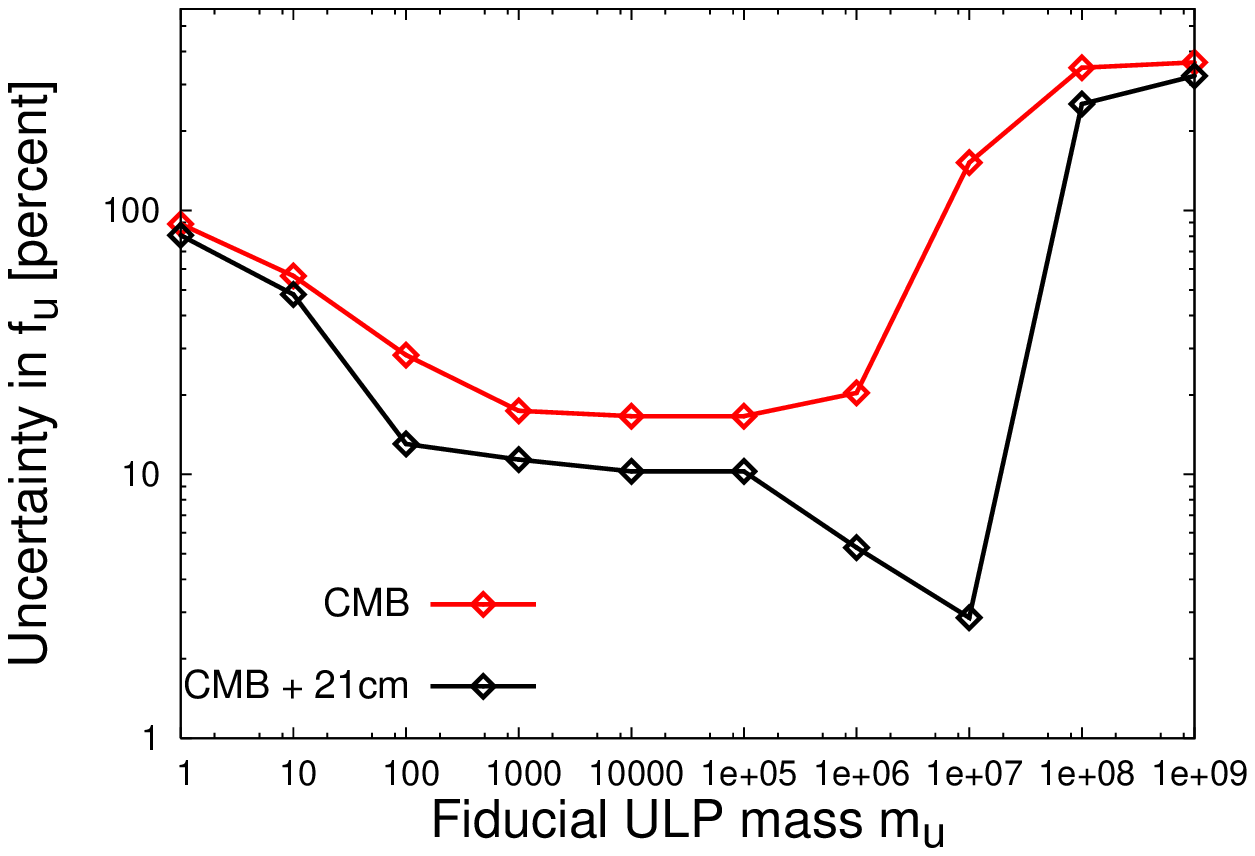}
  \end{center}
%  \caption{Uncertainty in fa}
%  \label{fig:one}
 \end{minipage}
 \begin{minipage}{0.5\hsize}
  \begin{center}
   \includegraphics[width=\linewidth]{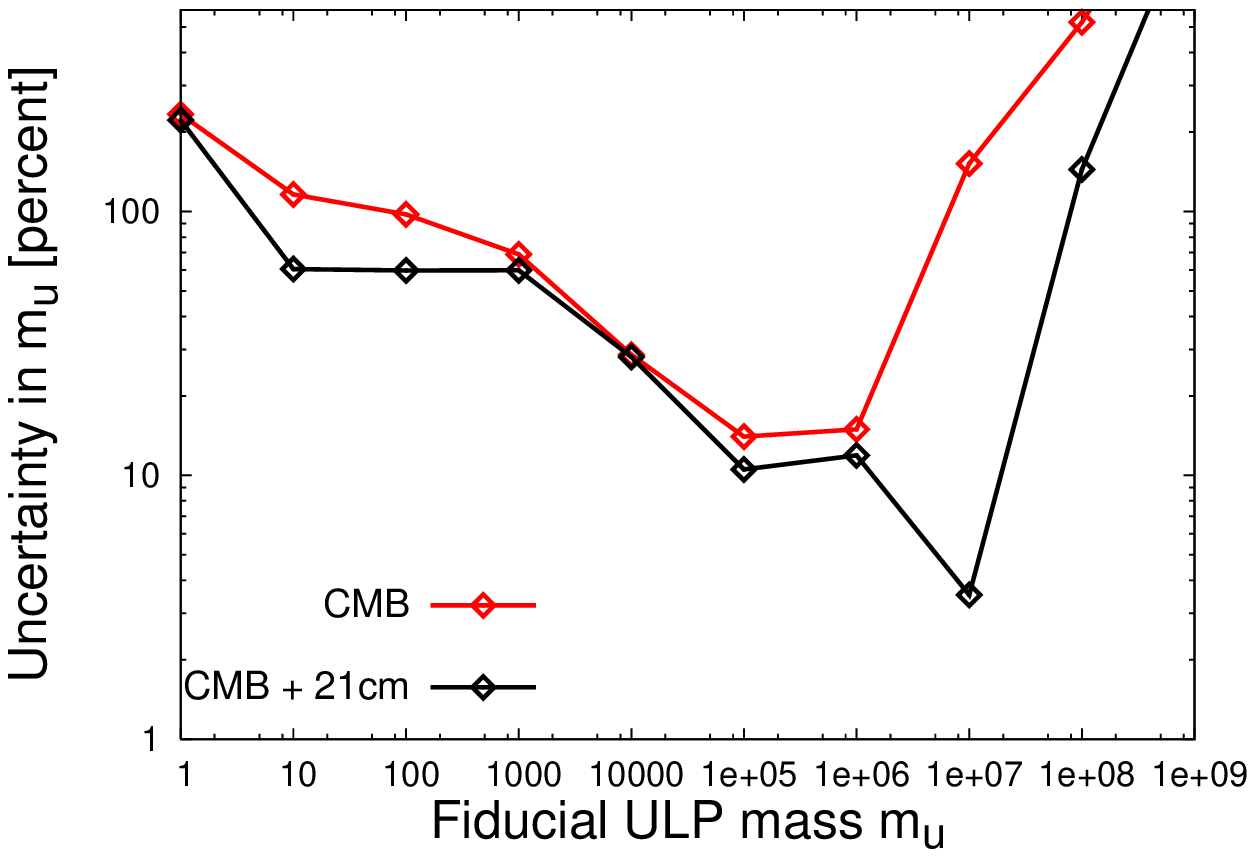}
  \end{center}
%  \caption{Uncertainty in ma}
%  \label{fig:two}
 \end{minipage}
\caption {$1 \sigma$ errors in $f_u$ and $m_u$ (the fiducial value $f_u=0.05$) for several fiducial values of $m_u$ in terms of $H_0(\approx 2 \times 10^{-33}$ eV).}
\label{sigmafa005}
\end{figure}

%%%%%%%%%%%%%%%%%%%%%%%%%%%%%%%%%

%%%%%%%%%%%%%%%%%%%%%%%%%%%%%%%%%%%%%%%%
\begin{figure}[htbp]
 \begin{minipage}{0.5\hsize}
% \begin{minipage}{0.48\textwidth}
  \begin{center}
   \includegraphics[width=\linewidth]{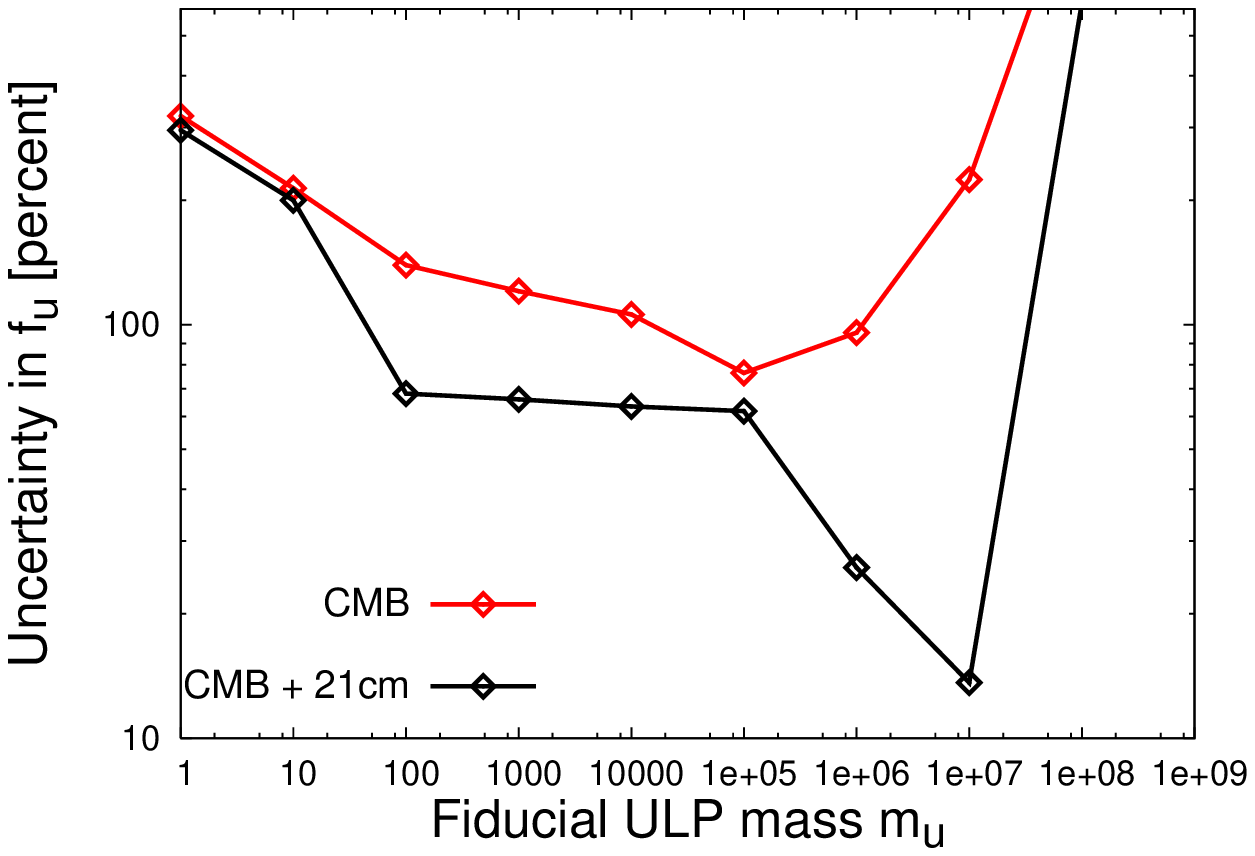}
  \end{center}
%  \caption{Uncertainty in fa}
%  \label{fig:one}
 \end{minipage}
 \begin{minipage}{0.5\hsize}
% \begin{minipage}{0.48\textwidth}
  \begin{center}
   \includegraphics[width=\linewidth]{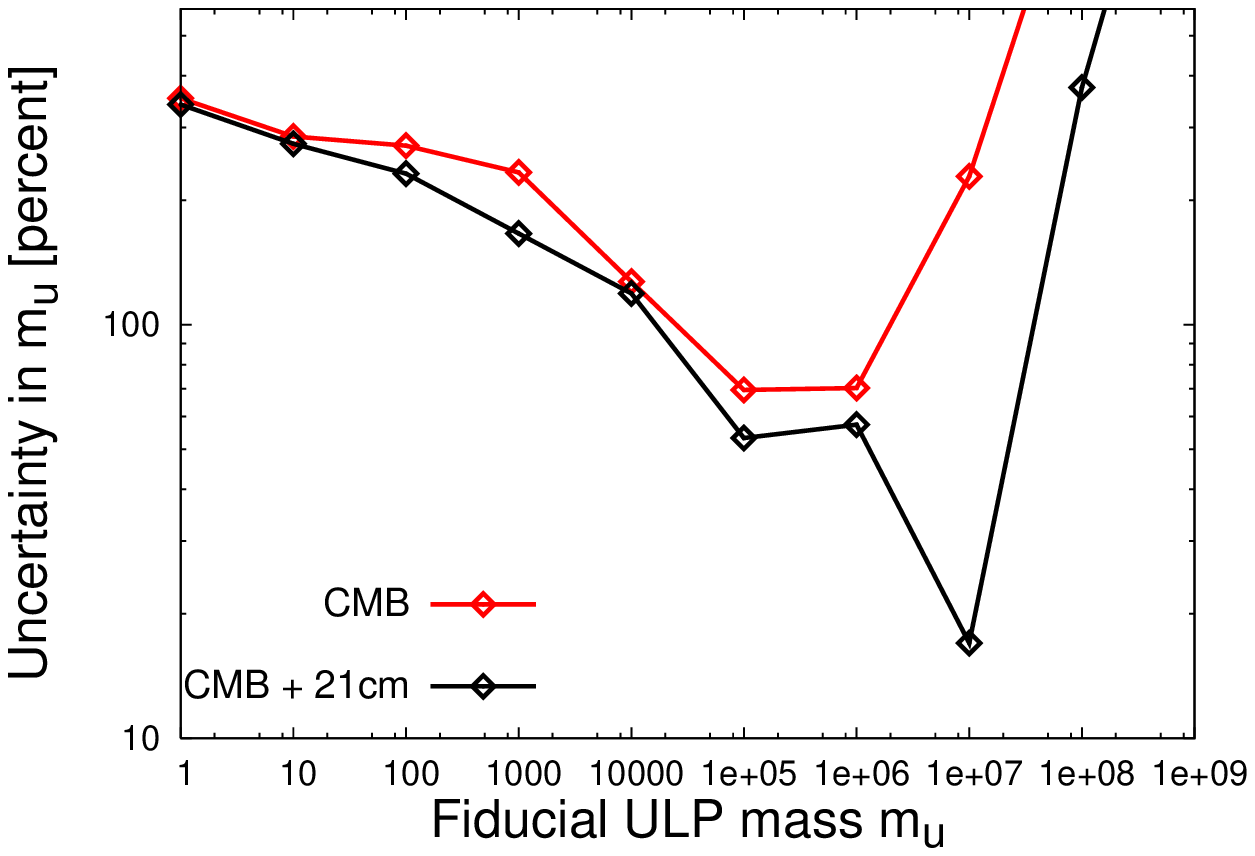}
  \end{center}
%  \caption{Uncertainty in ma}
%  \label{fig:two}
 \end{minipage}
\caption {$1 \sigma$ errors in $f_u$ and $m_u$ (the fiducial value $f_u=0.01$) for several fiducial values of $m_u$ in terms of $H_0(\approx 2 \times 10^{-33}$ eV).}
\label{sigmafa001}
\end{figure}

%%%%%%%%%%%%%%%%%%%%%%%%%%%%%%%%%

The main goal of this paper is to clarify the power of the 21 cm observables to constrain the ULP parameters, and our results are summarized in Fig. \ref{sigmafa005} which shows the $1\sigma$ uncertainties in the ULP parameters for several representative ULP masses for $f_u=0.05$. The $1\sigma$ errors on the ULP parameters $f_u,m_u$ can be of order a few percent for the mass range to which the 21 cm signals are most sensitive.  The sensitivity of the cosmological observables to the ULP parameters, however, depends on the fiducial values, and the errors for a smaller ULP fraction $f_u=0.01$ are shown in Fig. \ref{sigmafa001}. A bigger ULP fraction can imprint a bigger effect on the matter power, and hence a smaller error is forecasted as expected. Despite such quantitative changes in the error estimations, different ULP fraction cases share the common features: the 21 cm observables are most sensitive to the  ULP parameters when the ULP mass is around $m_u\sim 10^7 H_0$ which lets the ULPs start oscillations in the 21 cm observable range as inferred from the Fig. \ref{pertfig} showing the significant change in the matter power at the 21 cm observable scales $0.055 \lesssim k\lesssim  0.15$ [Mpc$^{-1}]$ (equivalently $0.08 h$/Mpc$ \lesssim k\lesssim  0.22 h/$Mpc). 
The sensitivity of the CMB observables to the ULPs increases on the other hand up to the ULP mass of about $10^5H_0$ which corresponds to the oscillation starting around the CMB last scattering epoch. For instance, we found numerically $2 \times 10^4 H_0 \sim H(z=1100)$ and we can indeed see that $\sigma(m_u)$ does not improve so much by adding the 21 cm observables for the mass around $m_u\sim 10^{4\sim 5}H_0$, which implies that the CMB constraint on $m_u$ is dominant over that from the 21 cm observables for this mass range. The CMB however starts losing its sensitivity to the ULPs significantly for the larger ULP masses $m_u \gtrsim 10^6 H_0$ which initiate the oscillations well before the last scattering epoch.

% The CMB is indeed sensitive to $m_u$  $m_u$ is quite   The 21 cm constraints on $f_u$ are in general more stringent than those on $m_u$. 

%% showing the matter power spectrum in the 21 am observable range $0.055 - 0.15 Mpc^{-1}$ adopted in our analysis.).
%% Both parameter uncertainties share the common features. 

%% The CMB is most sensitive to the ULP parameter when the ULP mass is around $m\sim 10^5 H_0$  which lets the ULPs start oscillation around the last scattering surface. 

%%  The axion mass parameters can be constrained to the order of $10\sim 20$ \% can be possible from the future 21 cm experiments. %The change in the matter radiation equality affecting the other CMB observables

%% the constraints from the 21cm observables becomes stringer up to around $m_u=10^7$ because the characteristic turn-over scale when the matter power spectrum suppression  falls in the 21cm observational range $0.055 - 0.15 Mpc^{-1}$. Consequently the power spectrum has the larger sensitivity to the variation of $f_u$ which can also be inferred from Fig. \ref{pertfig} showing the effects of $f_u$ on the power spectrum. For the ULP mass larger than $m_u$, the suppression scale goes beyond the 21cm sensitive scale and the constraints turn out to well exceed $100\%$. 

\section{Discussion/Conclusion}
We explored future prospects for setting constraints on ultra-light scalar fields from  21 cm observables. We found that the CMB including CMB lensing is most sensitive to the ULP mass range of $10^4 H_0\sim 10^6 H_0$ and the 21 cm is most sensitive to $m_u \sim 10^7 H_0$. We forecast that the future 21 cm can constrain the ULP parameters (the density fraction and the mass) with the order 10 \% accuracy (and even better with a few percent accuracy when the mass is around $m_u \sim 10^7 H_0$). Because of the complications due to  nonlinearity, however, the ULPs with $m_u \gg 10^7 H_0$ would be hard to probe by the large-scale structure of the Universe, even though these mass ranges can  be well probed by other probes such as  black holes and dwarf galaxies \cite{fuz,joe3,arv,tasi}. Further studies on the complimentarity between different observables are left for our future work.

\begin{figure}[!htb]
%\minipage{0.32\textwidth}
\minipage{0.48\textwidth}
  \includegraphics[width=\linewidth]{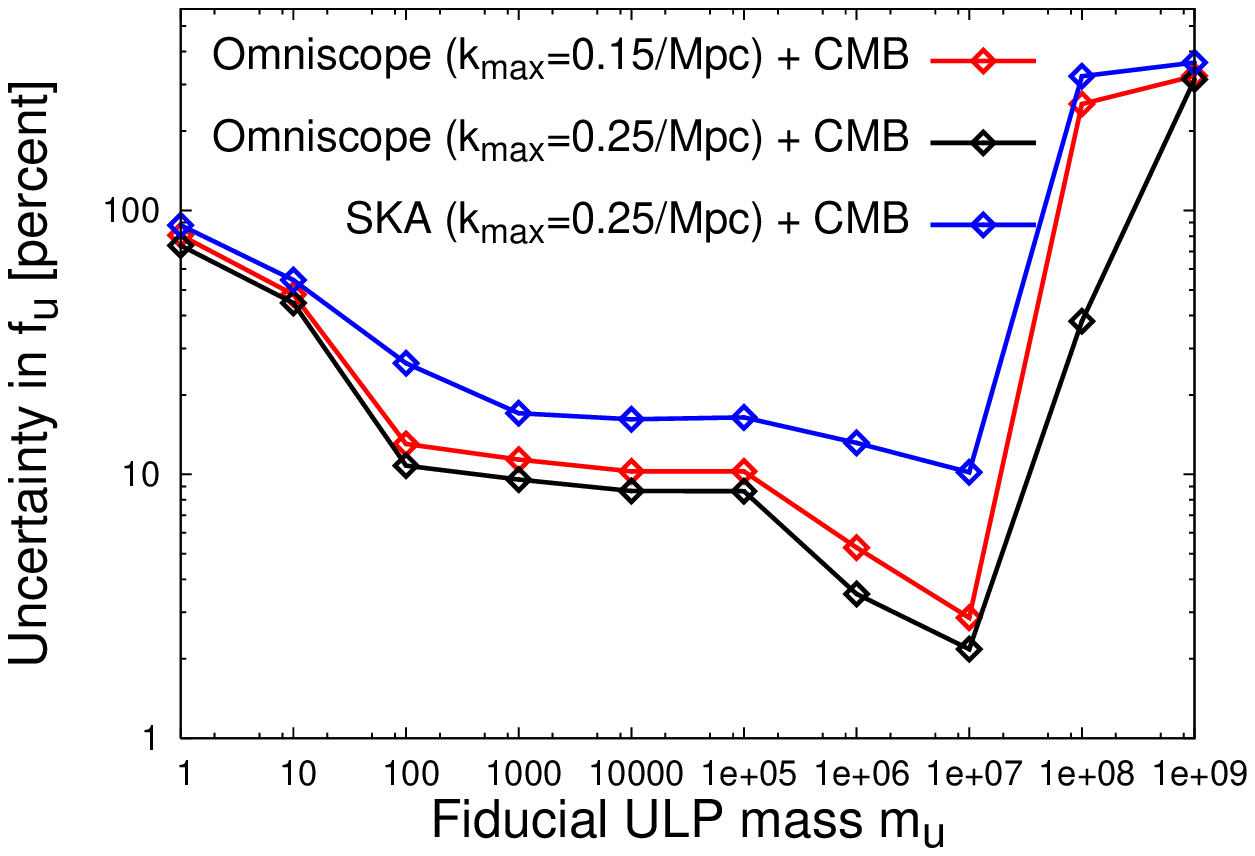}
%  \caption{A really Awesome Image} \label{fig:awesome_image1}
\endminipage\hfill
\minipage{0.48\textwidth}%
  \includegraphics[width=\linewidth]{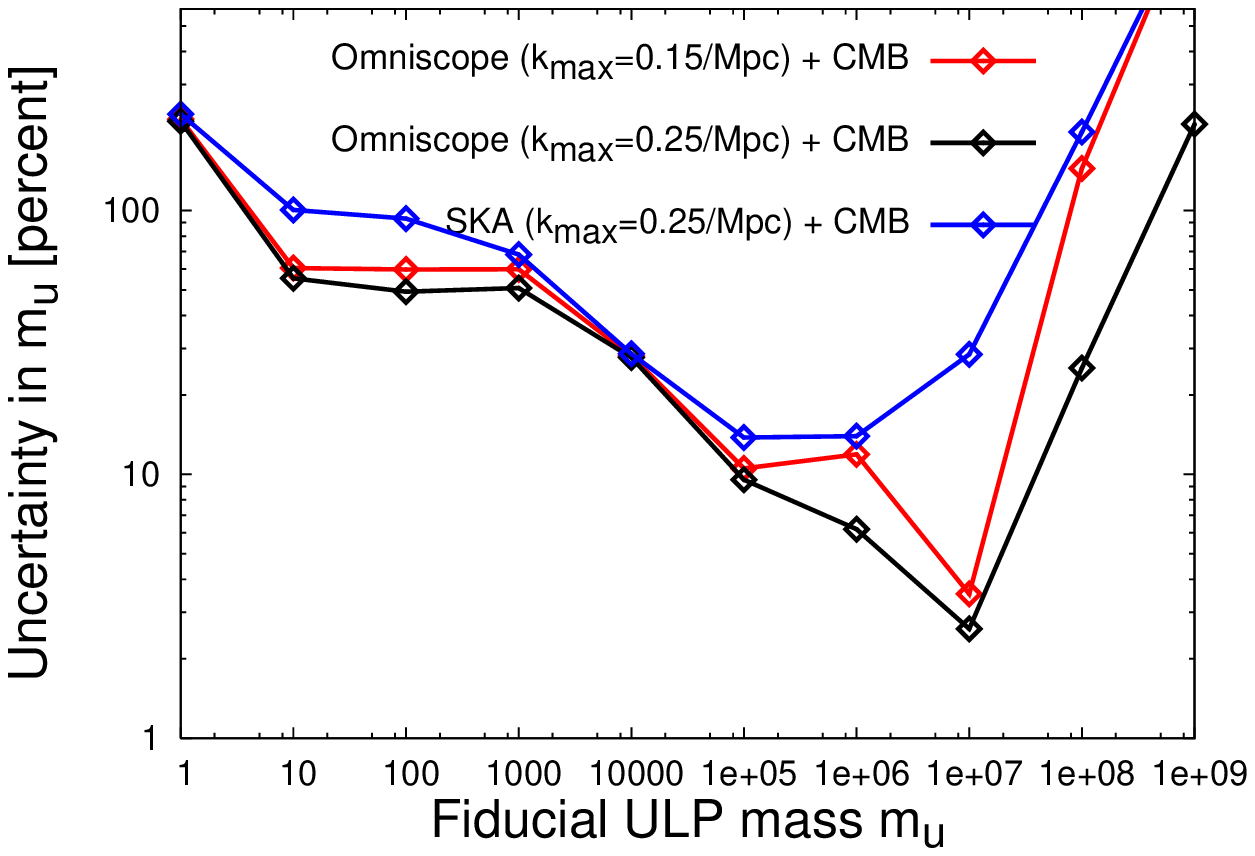}
%  \caption{A really Awesome Image}\label{fig:awesome_image3}
\endminipage
\caption{$1 \sigma$ error in $f_u$ and $m_u$ for different experiment specifications. The fiducial values of $m_u$ are in terms of $H_0(\approx 2 \times 10^{-33}$ eV).}
\label{compa}
\end{figure}

Before closing our discussions, let us briefly comment on the specification dependence for constraining the ULP parameters. Changing to a different redshift bin or changing the neutrino normal mass hierarchy to the inverted mass hierarchy pattern do not lead to an appreciable change in the ULP parameter bounds. A notable change however can result from changing $k_{max}$ to a bigger value which can be expected due to a larger number of available modes for a higher $k$. Fig. \ref{compa} shows the constraints on $f_u,m_u$ for $k_{max}=0.15$/Mpc and $0.25$/Mpc. Changing $k_{max}$ from our default value of 0.15/Mpc to 0.25/Mpc can easily improve the ULP constrains by $10 \%$ or more depending on the mass range. Higher values of $k$ also help to extend the 21 cm-sensitive scale to a smaller scale. We also demonstrated in Fig. \ref{compa}, how the error estimation can be affected by the experimental specifications, the constrains from the SKA-like experiment \cite{ska2} being specified by $(N_{in},L_{min},\eta,A_e(z=6/8/12)[m^2],\Omega[sr])=(1400, 10, 0.8, 30/50/104,\lambda^2/A_e)$ ($\lambda=21(1+z)$, and we assumed the same observation time and band width as those for the Omniscope telescope which has been assumed for the main body of the paper). The $k$ range well beyond the scales considered here will significantly be affected by nonlinearity, and the ULP constraints including those of small-scale physics not considered here such as inhomogeneous reionization and nonlinear bias would deserve further studies. Even for the SKA-like experiment, the 21 cm observables are still powerful enough to constrain $f_u$ with  order $10\%$ precision and also $m_u$ even though not as stringently as the constraints on $f_u$. The slight worsening of the bounds on $m_u$ around $m_u\sim 10^6 H_0$ is partly due to the CMB losing the sensitivity on $m_u$ even though it is overwhelmed by the sensitivity enhancement of the 21 cm signal for $m_u\sim 10^7 H_0$. We also note, for $m_u\sim 10^4 H_0$, the ULP oscillations start around the last scattering epoch, and the CMB observables overwhelm the 21 cm observables in constraining $m_u$.
%%%%%%%%%%%%%%%%%%%%%%%%%%%%%%%%%%%%%%%%%%%%%%%
%%%%%%%%%%%%%%%%%%%%%%%%%%%%%%%%%%%%%%%%%%%%%%%%%%%%%%%%%%

The experimental specifications to observe the 21 cm emission signals used in our analysis are sensitive to  quasi-linear scales, and we have used the matter power spectrum using  Halofit to take account of the nonlinearity. We first obtained the linear power spectra including the ULPs and neutrinos, and applied the Halofit formula to convert this linear power spectrum to the nonlinear power spectrum. In this mapping from the linear to the nonlinear power spectrum, the effects of the light species such as the ULPs and neutrinos are conventionally not included and this would be worth further exploration. Our studies can be extended to scenarios including  multiple ULPs with different masses, which is also left for  future work.

\section*{Acknowledgement}
We thank K. Choi and D. Marsh for the useful discussions. This work was supported by the MEXT of Japan and by French state funds managed by the ANR within the Investissements d'Avenir programme under reference ANR-11-IDEX-0004-02.  The research of JS has been supported at IAP by the ERC project 267117 (DARK) hosted by Universit\'e Pierre et Marie Curie - Paris 6 and at JHU by NSF grant OIA-1124403. K.K. thanks the hospitality of IAP where this work was initiated.
%We thank K. Choi and D. Marsh for the useful discussions. This work was supported by the MEXT of Japan and French state funds managed by the ANR.  K.K. thanks the hospitality of IAP where this work was initiated.
%%%%%%%%%%%%%%%%%%%%%%%%%%%%%%%%%%%%%%%


\begin{thebibliography}{99}


%\cite{Peccei:1977hh}
\bibitem{pec} 
  R.~D.~Peccei and H.~R.~Quinn,
  %``CP Conservation in the Presence of Instantons,''
  Phys.\ Rev.\ Lett.\  {\bf 38}, 1440 (1977).
  %%CITATION = PRLTA,38,1440;%%
  %2672 citations counted in INSPIRE as of 02 Dec 2013

%\cite{Weinberg:1977ma}
\bibitem{wein} 
  S.~Weinberg,
  %``A New Light Boson?,''
  Phys.\ Rev.\ Lett.\  {\bf 40}, 223 (1978).
  %%CITATION = PRLTA,40,223;%%
  %1892 citations counted in INSPIRE as of 02 Dec 2013

%\cite{Wilczek:1977pj}
\bibitem{wil} 
  F.~Wilczek,
  %``Problem of Strong p and t Invariance in the Presence of Instantons,''
  Phys.\ Rev.\ Lett.\  {\bf 40}, 279 (1978).
  %%CITATION = PRLTA,40,279;%%
  %1827 citations counted in INSPIRE as of 02 Dec 2013


%\cite{Arvanitaki:2009fg}
\bibitem{john1} 
  A.~Arvanitaki, S.~Dimopoulos, S.~Dubovsky, N.~Kaloper and J.~March-Russell,
  %``String Axiverse,''
  Phys.\ Rev.\ D {\bf 81}, 123530 (2010)
  [arXiv:0905.4720 [hep-th]].
  %%CITATION = ARXIV:0905.4720;%%

%\cite{}
\bibitem{ed} 
  E.~Witten,
  %``Some Properties of O(32) Superstrings,''
  Phys.\ Lett.\ B {\bf 149}, 351 (1984).
  %%CITATION = PHLTA,B149,351;%%
  %570 citations counted in INSPIRE as of 02 Dec 2013

%\cite{Svrcek:2006yi}
\bibitem{sv} 
  P.~Svrcek and E.~Witten,
  %``Axions In String Theory,''
  JHEP {\bf 0606}, 051 (2006)
  [hep-th/0605206].
  %%CITATION = HEP-TH/0605206;%%
  %237 citations counted in INSPIRE as of 22 Jan 2014

%\cite{Acharya:2010zx}
\bibitem{acha} 
  B.~S.~Acharya, K.~Bobkov and P.~Kumar,
  %``An M Theory Solution to the Strong CP Problem and Constraints on the Axiverse,''
  JHEP {\bf 1011}, 105 (2010)
  [arXiv:1004.5138 [hep-th]].
  %%CITATION = ARXIV:1004.5138;%%
  %45 citations counted in INSPIRE as of 02 Dec 2013


%\cite{Hu:2000ke}
\bibitem{fuz} 
  W.~Hu, R.~Barkana and A.~Gruzinov,
  %``Cold and fuzzy dark matter,''
  Phys.\ Rev.\ Lett.\  {\bf 85}, 1158 (2000)
  [astro-ph/0003365].
  %%CITATION = ASTRO-PH/0003365;%%
  %160 citations counted in INSPIRE as of 12 Apr 2013

%\cite{Sikivie:2009qn}
\bibitem{sik} 
  P.~Sikivie and Q.~Yang,
  %``Bose-Einstein Condensation of Dark Matter Axions,''
  Phys.\ Rev.\ Lett.\  {\bf 103}, 111301 (2009)
  [arXiv:0901.1106 [hep-ph]].
  %%CITATION = ARXIV:0901.1106;%%
  %80 citations counted in INSPIRE as of 10 Aug 2013

%\cite{Hu:1998kj}
\bibitem{wayne4} 
  W.~Hu,
  %``Structure formation with generalized dark matter,''
  Astrophys.\ J.\  {\bf 506}, 485 (1998)
  [astro-ph/9801234].
  %%CITATION = ASTRO-PH/9801234;%%
  %200 citations counted in INSPIRE as of 25 Apr 2013

%\cite{Kolb:1990vq}
\bibitem{kol} 
  E.~W.~Kolb and M.~S.~Turner, {\it The Early Universe}, Front.\ Phys.\  {\bf 69}, 1 (1990).
  %%CITATION = FRPHA,69,1;%%
  %588 citations counted in INSPIRE as of 20 Mar 2014

%\cite{Marsh:2013ywa}
\bibitem{joe3} 
  D.~J.~E.~Marsh and J.~Silk,
  %``A Model For Halo Formation With Axion Mixed Dark Matter,''
  arXiv:1307.1705 [astro-ph.CO].
  %%CITATION = ARXIV:1307.1705;%%
  %1 citations counted in INSPIRE as of 07 Aug 2013

%\cite{Barbieri:2005gj}
\bibitem{hall} 
  R.~Barbieri, L.~J.~Hall, S.~J.~Oliver and A.~Strumia,
  %``Dark energy and right-handed neutrinos,''
  Phys.\ Lett.\ B {\bf 625}, 189 (2005)
  [hep-ph/0505124].
  %%CITATION = HEP-PH/0505124;%%
  %41 citations counted in INSPIRE as of 06 May 2013


%\cite{Amendola:2005ad}
\bibitem{amen} 
  L.~Amendola and R.~Barbieri,
  %``Dark matter from an ultra-light pseudo-Goldsone-boson,''
  Phys.\ Lett.\ B {\bf 642}, 192 (2006)
  [hep-ph/0509257].
  %%CITATION = HEP-PH/0509257;%%
  %10 citations counted in INSPIRE as of 08 Apr 2013

%\cite{Abazajian:2011dt}
\bibitem{kev4} 
  K.~N.~Abazajian, E.~Calabrese, A.~Cooray, F.~De Bernardis, S.~Dodelson, A.~Friedland, G.~M.~Fuller and S.~Hannestad {\it et al.},
  %``Cosmological and Astrophysical Neutrino Mass Measurements,''
  Astropart.\ Phys.\  {\bf 35}, 177 (2011)
  [arXiv:1103.5083 [astro-ph.CO]].
  %%CITATION = ARXIV:1103.5083;%%
  %63 citations counted in INSPIRE as of 01 Dec 2013

%\cite{Lesgourgues:2005yv}
\bibitem{les} 
  J.~Lesgourgues, L.~Perotto, S.~Pastor and M.~Piat,
  %``Probing neutrino masses with cmb lensing extraction,''
  Phys.\ Rev.\ D {\bf 73}, 045021 (2006)
  [astro-ph/0511735].
  %%CITATION = ASTRO-PH/0511735;%%
  %83 citations counted in INSPIRE as of 31 Oct 2013


%\cite{}
\bibitem{wan} 
  X.~Wang, X.~-L.~Meng, T.~-J.~Zhang, H.~Shan, Y.~Gong, C.~Tao, X.~Chen and Y.~F.~Huang
  %``Observational constraints on cosmic neutrinos and dark energy revisited,''
  arXiv:1210.2136 [astro-ph.CO].
  %%CITATION = ARXIV:1210.2136;%%

%\cite{Saito:2008bp}
\bibitem{sai} 
  S.~Saito, M.~Takada and A.~Taruya,
  %``Impact of massive neutrinos on nonlinear matter power spectrum,''
  Phys.\ Rev.\ Lett.\  {\bf 100}, 191301 (2008)
  [arXiv:0801.0607 [astro-ph]].
  %%CITATION = ARXIV:0801.0607;%%
  %57 citations counted in INSPIRE as of 12 Nov 2013

%\cite{Bird:2011rb}
\bibitem{bir} 
  S.~Bird, M.~Viel and M.~G.~Haehnelt,
  %``Massive Neutrinos and the Non-linear Matter Power Spectrum,''
  Mon.\ Not.\ Roy.\ Astron.\ Soc.\  {\bf 420}, 2551 (2012)
  [arXiv:1109.4416 [astro-ph.CO]].
  %%CITATION = ARXIV:1109.4416;%%
  %30 citations counted in INSPIRE as of 12 Nov 2013


%\cite{Battye:2013xqa}
\bibitem{bat} 
  R.~A.~Battye and A.~Moss,
  %``Evidence for massive neutrinos from CMB and lensing observations,''
  arXiv:1308.5870 [astro-ph.CO].
  %%CITATION = ARXIV:1308.5870;%%
  %4 citations counted in INSPIRE as of 12 Nov 2013



%\cite{Takeuchi:2013gpa}
\bibitem{takeu} 
  Y.~Takeuchi and K.~Kadota,
  %``Probing Neutrinos from Planck and Forthcoming Galaxy Redshift Surveys,''
  arXiv:1310.0037 [astro-ph.CO].
  %%CITATION = ARXIV:1310.0037;%%


%\cite{Marsh:2010wq}
\bibitem{david3} 
  D.~J.~E.~Marsh and P.~G.~Ferreira,
  %``Ultra-Light Scalar Fields and the Growth of Structure in the Universe,''
  Phys.\ Rev.\ D {\bf 82}, 103528 (2010)
  [arXiv:1009.3501 [hep-ph]].
  %%CITATION = ARXIV:1009.3501;%%





%\cite{Arvanitaki:2010sy}
\bibitem{arv} 
  A.~Arvanitaki and S.~Dubovsky,
  %``Exploring the String Axiverse with Precision Black Hole Physics,''
  Phys.\ Rev.\ D {\bf 83}, 044026 (2011)
  [arXiv:1004.3558 [hep-th]].
  %%CITATION = ARXIV:1004.3558;%%
  %46 citations counted in INSPIRE as of 08 Aug 2013



%\cite{Ringwald:2013via}
\bibitem{rin} 
  A.~Ringwald,
  %``Ultralight Particle Dark Matter,''
  arXiv:1310.1256 [hep-ph].
  %%CITATION = ARXIV:1310.1256;%%

%\cite{Marsh:2011bf}
\bibitem{pedro} 
  D.~J.~E.~Marsh, E.~Macaulay, M.~Trebitsch and P.~G.~Ferreira,
  %``Ultra-light Axions: Degeneracies with Massive Neutrinos and Forecasts for Future Cosmological Observations,''
  Phys.\ Rev.\ D {\bf 85}, 103514 (2012)
  [arXiv:1110.0502 [astro-ph.CO]].
  %%CITATION = ARXIV:1110.0502;%%
  %8 citations counted in INSPIRE as of 08 Apr 2013


%\cite{Davidson:2013aba}
\bibitem{davi} 
  S.~Davidson and M.~Elmer,
  %``Bose Einstein condensation of the classical axion field in cosmology?,''
  arXiv:1307.8024 [hep-ph].
  %%CITATION = ARXIV:1307.8024;%%

%\cite{Graham:2013gfa}
\bibitem{gra} 
  P.~W.~Graham and S.~Rajendran,
  %``New Observables for Direct Detection of Axion Dark Matter,''
  Physical Review D 88, {\bf 035023} (2013)
  [Phys.\ Rev.\ D {\bf 88}, 035023 (2013)]
  [arXiv:1306.6088 [hep-ph]].
  %%CITATION = ARXIV:1306.6088;%%
  %2 citations counted in INSPIRE as of 25 Sep 2013


%\cite{Lewis:2006fu}
\bibitem{lewrep} 
  A.~Lewis and A.~Challinor,
  %``Weak gravitational lensing of the cmb,''
  Phys.\ Rept.\  {\bf 429}, 1 (2006)
  [astro-ph/0601594].
  %%CITATION = ASTRO-PH/0601594;%%


%\cite{Furlanetto:2006jb}
\bibitem{furrep} 
  S.~Furlanetto, S.~P.~Oh and F.~Briggs,
  %``Cosmology at Low Frequencies: The 21 cm Transition and the High-Redshift Universe,''
  Phys.\ Rept.\  {\bf 433}, 181 (2006)
  [astro-ph/0608032].
  %%CITATION = ASTRO-PH/0608032;%%
  %246 citations counted in INSPIRE as of 13 Jun 2013







%\cite{Bond:1980ha}
\bibitem{bond} 
  J.~R.~Bond, G.~Efstathiou and J.~Silk,
  %``Massive Neutrinos and the Large Scale Structure of the Universe,''
  Phys.\ Rev.\ Lett.\  {\bf 45}, 1980 (1980).
  %%CITATION = PRLTA,45,1980;%%
  %262 citations counted in INSPIRE as of 07 Aug 2013

%\cite{Lewis:1999bs}
\bibitem{camb} 
  A.~Lewis, A.~Challinor, A.~Lasenby
  %``Efficient computation of CMB anisotropies in closed FRW models,''
  Astrophys.\ J.\  {\bf 538}, 473 (2000)
  [astro-ph/9911177].
  %%CITATION = ASTRO-PH/9911177;%%
  %1125 citations counted in INSPIRE as of 28 Mar 2013


\bibitem{smi} 
  R.~E.~Smith {\it et al.}  [Virgo Consortium Collaboration],
  %``Stable clustering, the halo model and nonlinear cosmological power spectra,''
  Mon.\ Not.\ Roy.\ Astron.\ Soc.\  {\bf 341}, 1311 (2003)
  [astro-ph/0207664].
  %%CITATION = ASTRO-PH/0207664;%%
  %797 citations counted in INSPIRE as of 12 Nov 2013


%\cite{Takahashi:2012em}
\bibitem{ryo} 
  R.~Takahashi, M.~Sato, T.~Nishimichi, A.~Taruya and M.~Oguri,
  %``Revising the Halofit Model for the Nonlinear Matter Power Spectrum,''
  Astrophys.\ J.\  {\bf 761}, 152 (2012)
  [arXiv:1208.2701 [astro-ph.CO]].
  %%CITATION = ARXIV:1208.2701;%%
  %17 citations counted in INSPIRE as of 12 Nov 2013

%\cite{Meiksin:1998ra}
\bibitem{mei} 
  A.~Meiksin, M.~J.~White and J.~A.~Peacock,
  %``Baryonic signatures in large scale structure,''
  Mon.\ Not.\ Roy.\ Astron.\ Soc.\  {\bf 304}, 851 (1999)
  [astro-ph/9812214].
  %%CITATION = ASTRO-PH/9812214;%%
  %132 citations counted in INSPIRE as of 05 Dec 2013

%\cite{Crocce:2007dt}
\bibitem{cro} 
  M.~Crocce and R.~Scoccimarro,
  %``Nonlinear Evolution of Baryon Acoustic Oscillations,''
  Phys.\ Rev.\ D {\bf 77}, 023533 (2008)
  [arXiv:0704.2783 [astro-ph]].
  %%CITATION = ARXIV:0704.2783;%%
  %183 citations counted in INSPIRE as of 05 Dec 2013

%\cite{Seo:2009fp}
\bibitem{seo} 
  H.~-J.~Seo, J.~Eckel, D.~J.~Eisenstein, K.~Mehta, M.~Metchnik, N.~Padmanabhan, P.~Pinto and R.~Takahashi {\it et al.},
  %``High-precision predictions for the acoustic scale in the non-linear regime,''
  Astrophys.\ J.\  {\bf 720}, 1650 (2010)
  [arXiv:0910.5005 [astro-ph.CO]].
  %%CITATION = ARXIV:0910.5005;%%
  %37 citations counted in INSPIRE as of 05 Dec 2013

%\cite{Taruya:2009ir}
\bibitem{taruya4} 
  A.~Taruya, T.~Nishimichi, S.~Saito and T.~Hiramatsu,
  %``Non-linear Evolution of Baryon Acoustic Oscillations from Improved Perturbation Theory in Real and Redshift Spaces,''
  Phys.\ Rev.\ D {\bf 80}, 123503 (2009)
  [arXiv:0906.0507 [astro-ph.CO]].
  %%CITATION = ARXIV:0906.0507;%%
  %68 citations counted in INSPIRE as of 05 Dec 2013


%\cite{Jeong:2006xd}
\bibitem{jeon} 
  D.~Jeong and E.~Komatsu,
  %``Perturbation theory reloaded: analytical calculation of non-linearity in baryonic oscillations in the real space matter power spectrum,''
  Astrophys.\ J.\  {\bf 651}, 619 (2006)
  [astro-ph/0604075].
  %%CITATION = ASTRO-PH/0604075;%%
  %100 citations counted in INSPIRE as of 05 Dec 2013

%\cite{Eisenstein:1997jh}
\bibitem{eh97} 
  D.~J.~Eisenstein and W.~Hu,
  %``Power spectra for cold dark matter and its variants,''
  Astrophys.\ J.\  {\bf 511}, 5 (1997)
  [astro-ph/9710252].
  %%CITATION = ASTRO-PH/9710252;%%


%% \bibitem{D'Aloisio13}
%% A.~D'Aloisio, J.~Zhang, P.~R.~Shapiro, and Y.~Mao, MNRAS, {\bf 433}, 2900 (2013) [arXiv:1304.6411[astro-ph.CO]].
 
%% \bibitem{Mao13}
%%  Y.~Mao, A.~D'Aloisio, J.~Zhang, and P.~R.~Shapiro, Phys.\ Rev.\ D {\bf 88}, 081303 (2013) [arXiv:1305.0313[astro-ph.CO]].

%\cite{Furlanetto:2004nh}
\bibitem{esmr} 
  S.~Furlanetto, M.~Zaldarriaga and L.~Hernquist,
  %``The Growth of HII regions during reionization,''
  Astrophys.\ J.\  {\bf 613}, 1 (2004)
  [astro-ph/0403697].
  %%CITATION = ASTRO-PH/0403697;%%
  %222 citations counted in INSPIRE as of 28 Nov 2013






%\cite{McQuinn:2005hk}
\bibitem{mc} 
  M.~McQuinn, O.~Zahn, M.~Zaldarriaga, L.~Hernquist and S.~R.~Furlanetto,
  %``Cosmological parameter estimation using 21 cm radiation from the epoch of reionization,''
  Astrophys.\ J.\  {\bf 653}, 815 (2006)
  [astro-ph/0512263].
  %%CITATION = ASTRO-PH/0512263;%%
  %171 citations counted in INSPIRE as of 28 Nov 2013

%\cite{Mao:2008ug}
\bibitem{mao} 
  Y.~Mao, M.~Tegmark, M.~McQuinn, M.~Zaldarriaga and O.~Zahn,
  %``How accurately can 21 cm tomography constrain cosmology?,''
  Phys.\ Rev.\ D {\bf 78}, 023529 (2008)
  [arXiv:0802.1710 [astro-ph]].
  %%CITATION = ARXIV:0802.1710;%%
  %72 citations counted in INSPIRE as of 12 Jun 2013



%\cite{Wyithe:2007if}
\bibitem{sys} 
  S.~Wyithe and M.~F.~Morales,
  %``Biased Reionisation and Non-Gaussianity in Redshifted 21cm Intensity Maps of the Reionisation Epoch,''
  %Submitted to: Mon.Not.Roy.Astron.Soc.
  [astro-ph/0703070 [ASTRO-PH]].
  %%CITATION = ASTRO-PH/0703070;%%
  %9 citations counted in INSPIRE as of 28 Nov 2013


%\cite{Tegmark:2008au}
\bibitem{teg1} 
%\cite{Tegmark:2009kv}
%\bibitem{Tegmark:2009kv} 
  M.~Tegmark and M.~Zaldarriaga,
  %``Omniscopes: Large Area Telescope Arrays with only N log N Computational Cost,''
  Phys.\ Rev.\ D {\bf 82}, 103501 (2010)
  [arXiv:0909.0001 [astro-ph.CO]], 
  %%CITATION = ARXIV:0909.0001;%%
  %20 citations counted in INSPIRE as of 01 Dec 2013
  M.~Tegmark and M.~Zaldarriaga,
  %``The Fast Fourier Transform Telescope,''
  Phys.\ Rev.\ D {\bf 79}, 083530 (2009)
  [arXiv:0805.4414 [astro-ph]].
  %%CITATION = ARXIV:0805.4414;%%
  %37 citations counted in INSPIRE as of 01 Dec 2013


%\cite{Shapiro:2012mq}
\bibitem{Shapiro13} 
  P.~R.~Shapiro, Y.~Mao, I.~T.~Iliev, G.~Mellema, K.~K.~Datta, K.~Ahn and J.~Koda,
  %``Will Nonlinear Peculiar Velocity and Inhomogeneous Reionization Spoil 21cm Cosmology from the Epoch of Reionization?,''
  Phys.\ Rev.\ Lett.\  {\bf 110}, 151301 (2013)
  [arXiv:1211.2036 [astro-ph.CO]].
  %%CITATION = ARXIV:1211.2036;%%
  %3 citations counted in INSPIRE as of 04 Dec 2013

%\cite{Bond:1997wr}
\bibitem{dick3} 
  J.~R.~Bond, G.~Efstathiou and M.~Tegmark,
  %``Forecasting cosmic parameter errors from microwave background anisotropy experiments,''
  Mon.\ Not.\ Roy.\ Astron.\ Soc.\  {\bf 291}, L33 (1997)
  [astro-ph/9702100].
  %%CITATION = ASTRO-PH/9702100;%%
  %475 citations counted in INSPIRE as of 03 Dec 2013

%\cite{Zaldarriaga:1997ch}
\bibitem{mati} 
  M.~Zaldarriaga, D.~N.~Spergel and U.~Seljak,
  %``Microwave background constraints on cosmological parameters,''
  Astrophys.\ J.\  {\bf 488}, 1 (1997)
  [astro-ph/9702157].
  %%CITATION = ASTRO-PH/9702157;%%
  %362 citations counted in INSPIRE as of 03 Dec 2013

%\cite{Sherwin:2011gv}
\bibitem{she} 
  B.~D.~Sherwin, J.~Dunkley, S.~Das, J.~W.~Appel, J.~R.~Bond, C.~S.~Carvalho, M.~J.~Devlin and R.~Dunner {\it et al.},
  %``Evidence for dark energy from the cosmic microwave background alone using the Atacama Cosmology Telescope lensing measurements,''
  Phys.\ Rev.\ Lett.\  {\bf 107}, 021302 (2011)
  [arXiv:1105.0419 [astro-ph.CO]].
  %%CITATION = ARXIV:1105.0419;%%
  %40 citations counted in INSPIRE as of 03 Dec 2013


%\cite{Howlett:2012mh}
\bibitem{how} 
  C.~Howlett, A.~Lewis, A.~Hall, A.~Challinor and ,
  %``CMB power spectrum parameter degeneracies in the era of precision cosmology,''
  JCAP {\bf 1204}, 027 (2012)
  [arXiv:1201.3654 [astro-ph.CO]].
  %%CITATION = ARXIV:1201.3654;%%
  %4 citations counted in INSPIRE as of 28 Mar 2013



%\cite{Ade:2013ktc}
\bibitem{adepl2} 
  P.~A.~R.~Ade {\it et al.}  [Planck Collaboration],
  %``Planck 2013 results. I. Overview of products and scientific results,''
  arXiv:1303.5062 [astro-ph.CO].
  %%CITATION = ARXIV:1303.5062;%%
  %200 citations counted in INSPIRE as of 03 Dec 2013


%\cite{Knox:1995dq}
\bibitem{knox} 
  L.~Knox,
  %``Determination of inflationary observables by cosmic microwave background anisotropy experiments,''
  Phys.\ Rev.\ D {\bf 52}, 4307 (1995)
  [astro-ph/9504054].
  %%CITATION = ASTRO-PH/9504054;%%
  %294 citations counted in INSPIRE as of 05 Dec 2013

%\cite{Hu:2001kj}
\bibitem{take} 
  W.~Hu and T.~Okamoto,
  %``Mass reconstruction with cmb polarization,''
  Astrophys.\ J.\  {\bf 574}, 566 (2002)
  [astro-ph/0111606].
  %%CITATION = ASTRO-PH/0111606;%%

%\cite{Okamoto:2003zw}
\bibitem{oka2} 
  T.~Okamoto and W.~Hu,
  %``CMB lensing reconstruction on the full sky,''
  Phys.\ Rev.\ D {\bf 67}, 083002 (2003)
  [astro-ph/0301031].
  %%CITATION = ASTRO-PH/0301031;%%


%\cite{Tegmark:1996bz}
\bibitem{kar} 
  M.~Tegmark, A.~Taylor and A.~Heavens
  %``Karhunen-Loeve eigenvalue problems in cosmology: How should we tackle large data sets?,''
  Astrophys.\ J.\  {\bf 480}, 22 (1997)
  [astro-ph/9603021].
  %%CITATION = ASTRO-PH/9603021;%%
  %375 citations counted in INSPIRE as of 30 Mar 2013


%\cite{Alvarez:2005sa}
\bibitem{koma} 
  M.~A.~Alvarez, E.~Komatsu, O.~Dore and P.~R.~Shapiro,
  %``The cosmic reionization history as revealed by the cmb doppler-21-cm correlation,''
  Astrophys.\ J.\  {\bf 647}, 840 (2006)
  [astro-ph/0512010].
  %%CITATION = ASTRO-PH/0512010;%%
  %31 citations counted in INSPIRE as of 16 Jun 2013







%\cite{Ade:2013zuv}
\bibitem{adepl} 
  P.~A.~R.~Ade {\it et al.}  [Planck Collaboration],
  %``Planck 2013 results. XVI. Cosmological parameters,''
  arXiv:1303.5076 [astro-ph.CO].
  %%CITATION = ARXIV:1303.5076;%%
  %953 citations counted in INSPIRE as of 02 Dec 2013



%\cite{Fogli:2012ua}
\bibitem{fogi3} 
  G.~L.~Fogli, E.~Lisi, A.~Marrone, D.~Montanino, A.~Palazzo, A.~M.~Rotunno,
  %``Global analysis of neutrino masses, mixings and phases: entering the era of leptonic CP violation searches,''
  Phys.\ Rev.\ D {\bf 86}, 013012 (2012)
  [arXiv:1205.5254 [hep-ph]].
  %%CITATION = ARXIV:1205.5254;%%
  %177 citations counted in INSPIRE as of 28 Mar 2013

%\cite{Beringer:1900zz}
\bibitem{pdg}
J.~Beringer {\it et al.}  [Particle Data Group Collaboration],
  %``Review of Particle Physics (RPP),''
  Phys.\ Rev.\ D {\bf 86}, 010001 (2012).
  %%CITATION = PHRVA,D86,010001;%%
  %1109 citations counted in INSPIRE as of 01 Apr 2013


%\cite{Tashiro:2013yea}
\bibitem{tasi} 
  H.~Tashiro, J.~Silk and D.~J.~E.~Marsh,
  %``Constraints on primordial magnetic fields from CMB distortions in the axiverse,''
  arXiv:1308.0314 [astro-ph.CO].
  %%CITATION = ARXIV:1308.0314;%%

\bibitem{ska2} 
http://www.skatelescope.org/

\end{thebibliography}
\end{document}